\documentstyle[epsfig]{mn}
\begin{document}
\def\ltsima{$\; \buildrel < \over \sim \;$}
\def\simlt{\lower.5ex\hbox{\ltsima}}
\def\gtsima{$\; \buildrel > \over \sim \;$}
\def\simgt{\lower.5ex\hbox{\gtsima}}
\def\approxgt{\mathrel{\hbox{\rlap{\lower.55ex \hbox {$\sim$}}
        \kern-.3em \raise.4ex \hbox{$>$}}}}
\def\approxlt{\mathrel{\hbox{\rlap{\lower.55ex \hbox {$\sim$}}
        \kern-.3em \raise.4ex \hbox{$<$}}}}

\title{First high-resolution detection of a warm absorber in the Broad Line Radio Galaxy 3C~382}

\author[E. Torresi et al.]
{E. Torresi$^{1,2}$, P. Grandi$^1$, A.L. Longinotti$^3$, M. Guainazzi$^4$, G.G.C. Palumbo$^2$,
\newauthor F. Tombesi$^2$, A. Nucita$^4$\\
$^1$Istituto di Astrofisica Spaziale e Fisica Cosmica-Bologna, INAF, 
via Gobetti 101, I-40129 Bologna \\
$^2$Dipartimento di Astronomia, Universit\`a di Bologna, via Ranzani 1, I-40127 Bologna, Italy \\
$^3$MIT Kavli Institute for Astrophysics and Space Research
77 Massachusetts Avenue, NE80-6011
Cambridge, MA 02139\\
$^4$European Space Astronomy Centre of ESA P.O.Box 78, Villanueva de la Canada, E-28691 Madrid, Spain
}

\maketitle

\begin{abstract}

Recent high-resolution measurements suggest that the soft X-ray spectrum of obscured Radio Galaxies (RG)  exhibits signatures of photoionised gas  (e.g. 3C~445 and 3C~33) similar to those observed in radio-quiet obscured Active Galactic Nuclei (AGN).
While signatures of warm absorbing gas covering a wide range of temperature and ionisation states have been detected in about one half of the population of nearby Seyfert 1 galaxies, no traces of warm absorber gas have been reported to date in the high-resolution spectra of Broad Line Radio Galaxies (BLRG). 
We present here the first detection of a soft X-ray warm absorber in the powerful FRII BLRG 3C~382 using the Reflection Grating Spectrometer (RGS) on-board \emph{XMM-Newton}. The absorption gas appears to be highly ionised, with column density of the order of 10$^{22}$ cm$^{-2}$,  ionisation parameter log$\xi>$2 erg~cm~s$^{-1}$ and  outflow velocities  of the order of 10$^{3}$ km s$^{-1}$.  The absorption lines may come from regions located outside the torus, however at distances less than 60 pc. This result may indicate that a plasma ejected at  velocities near the speed of light and a photoionised gas with slower outflow velocities can coexist in the same source beyond the Broad Line Regions.

\end{abstract}

\begin{keywords}
galaxies:active -- X-rays:galaxies -- warm absorber: individual: 3C~382
\end{keywords}

\vspace{1.0cm}

\section{Introduction}
The launch of \emph{XMM-Newton} and  \emph{Chandra} both carrying on-board diffraction grating spectrometers with high resolving power,  dramatically improved our knowledge of the gaseous environment of  Active Galactic Nuclei. At first, very bright and nearby Seyfert galaxies were explored and many results obtained in the last decade. Depending on the observer's line-of-sight, different features could be detected.  In X-ray obscured Seyfert 2s high-resolution observations have revealed soft X-ray spectra dominated by line emission from H-like and He-like elements as Neon, Oxygen, Carbon, as well as from Fe L-shell transitions (Fe XVII to Fe XXIV) (Sako et al. 2000; Kinkhabwala et al. 2002; Bianchi et al. 2005; Pounds \& Vaughan 2006; Guainazzi \& Bianchi 2007). The soft X-ray emission is most likely produced by photoionisation in a gas located in the Narrow Line Region (NLR) (Bianchi, Guainazzi \& Chiaberge 2006). 

Signatures of outflowing ionised gas have been detected in the X-ray spectra of about 50$\%$ of nearby Seyfert~1s. The data indicate the presence of a multi-phase gas with an ionisation parameter $\xi$ \footnote{$\xi$=$\displaystyle\frac{L}{n_{e}R^{2}}$, \emph{L} is the 1-1000 Rydberg source luminosity, \emph{n$_{e}$} is the gas density and \emph{R} is the distance from the source (Tarter Tucker and Salpeter 1969)}  spanning three orders of magnitude and column densities ranging between 10$^{20-22}$~cm$^{-2}$, with velocities of the order of a few hundreds to thousands km~s$^{-1}$ (Blustin et al. 2005; McKernan et al. 2007). Where these winds originate and are powered is still an open question. Several regions at different distances from the central engine, like  the NLR, the obscuring torus and the accretion disk have been proposed as a possible location for the gas (Crenshaw et al. 2003 and references therein).

For Radio Galaxies the investigation of the nuclear environment through high-resolution spectroscopy has just started. There are some hints that in obscured RGs (seen at high inclination angles) the soft X-ray spectra are similar to their Radio-Quiet (RQ) counterparts, Seyfert 2s. Evidence for photoionised gas has already been revealed by emission lines in the soft X-ray spectra of 3C 445 (Grandi et al. 2007; Sambruna et al. 2007) and 3C 33 (Torresi et al. 2009). In the latter case, the morphological coincidence between soft X-ray and optical [OIII]$\lambda$5007 emission suggests the NLR as the location of such gas (Torresi et al. 2009).

The current picture is less clear for the radio-loud (RL) counterparts of Seyfert 1s, where Doppler amplification of the jet flux can play a key role in blurring the spectral features: for example, RGs are known to have weaker Compton reflection signatures and weaker Fe K$\alpha$ lines when compared to Seyfert 1s (Grandi, Malaguti \& Fiocchi 2006 and references therein). Up to now, the search for warm absorbers in unobscured RGs (seen at low inclination angles) has been unsuccessful, except for a claim of its existence in 3C~382 and 3C~390.3 from analysis of ASCA data (Reynolds 1997). Therefore it is completely unclear whether winds can form and/or survive in the nuclear regions of RL AGNs, where large amounts of the gravitational power is funneled into relativistic collimated plasma. Addressing this point is fundamental to clarify the physical link among accretion flows, winds and jet, and the influence that each of these components can have on the host galaxy.

3C~382 is a nearby Radio Galaxy (z=0.0579), with lobe-dominated Fanaroff--Riley II radio morphology. It exhibits a long jet extending north--east of the core for 1.68' and two radio lobes, with a total extension of 3' (Black et al. 1992).\\
At optical wavelengths 3C~382 shows strong and broad lines (FWZI$>$25,000 km~s$^{-1}$), which are variable on timescales of months to years.  An \emph{Hubble Space Telescope (HST)} WFPC2 image showed that this source is an elliptical galaxy strongly dominated by an unresolved nucleus (Martel et al. 1999). The optical emission of this source is clearly photoionised as can be inferred from the de-reddened line ratios, i.e. log[OIII]/H$\beta$=0.90 and log[NII]/H$\alpha$=0.17 (Buttiglione et al. 2009), compared with the diagnostic diagrams of Miller et al. (2003).\\
In the X--ray band 3C~382 is a bright source (F$_{2-10 keV } \sim$ 3$\times$10$^{-11}$~erg~cm$^{-2}$~s$^{-1}$, from this \emph{XMM-Newton} observation). The 2--10 keV spectrum is well fitted with a single power-law; when the fit is extrapolated to lower energies a strong soft-excess is observed (Prieto 2000; Grandi et al. 2001). ROSAT/HRI observations revealed extended X-ray (0.2-2.4 keV) emission around 3C~382 (Prieto 2000), that was confirmed by the spatial resolved study of Gliozzi et al. (2007). However, as discussed by Grandi et al. (2001) and Gliozzi et al. (2007) the extended thermal emission cannot account for the observed soft-excess, which must be instead of nuclear origin.\\
Here we report the first high-resolution detection of an X-ray warm absorber in this BLRG.

\section{Observations and data reduction}
3C~382 was observed by \emph{XMM-Newton} on 8th April 2008 (OBSID: 0506120101) for about 40 ks. The data from EPIC (Str\"uder et al. 2001), RGS (den Herder et al.  2001) and Optical Monitor (OM, Mason et al. 2001) instruments were operating. As we are primarily interested in the warm absorber features, only the RGS data (6--38 $ \AA$)  are presented in this paper. Data were processed with the task \emph{rgsproc} of the SAS 8.0.0, which combines the event lists from all RGS CCD, produces source and background spectra using standard extraction regions, and generates response matrices. The background is estimated taking into account events from a region spatially offset from the source. 
All spectral fits presented in this work include absorption due to a line-of-sight Galactic column density of N$_{H}$=7.0$\times 10^{20}$ cm$^{-2}$ (Kalberla et al. 2005).
The cosmological parameters used throughout the paper are  $H_0=70$ km~s$^{-1}$~Mpc$^{-1}$, $\Omega_{m}=0.3$, $\Omega_{\Lambda} = 0.7$ (Spergel et al. 2007). 
The spectral analysis was performed using the \emph{XSPEC} v.11.3.2 package (Arnaud 1996). 

\begin{table*}
\centering
\caption{Soft X-ray absorption and  emission lines detected in the RGS spectrum of 3C~382.  Columns: (1) Identification of the transition; (2) 3C~382 rest-frame wavelength (in $\AA$); (3) Laboratory wavelength (in $\AA$); (4) Emission-line intensity; (5) Equivalent width (in m$\AA$); (6) $\Delta$C.}
\label{tab1}      
\centering          
\begin{tabular}{c c c c c c}     
\\
\hline\hline\                    
         
Transition             &$\lambda$(rest)            &$\lambda$(lab)               &Flux                                                         &EW$^{(a)}$         &$\Delta$C $^{(b)}$\\                                                                                                                               
-                                    & ($\AA$)                                 &($\AA$)                            &(10$^{-5}$photons cm$^{-2}$ s$^{-1}$)    &(m$\AA$)               &  -   \\                           
\hline
\\
NeX                    &12.07                                      &12.13                       &    -                                          &67$^{+34}_{-30}$                                  &$\Delta$C=7 \\
\\
FeXX                   &     12.78                                         &12.82                  &    -            &43$^{+31}_{-30}$       &$\Delta$C=8 \\
\\
OVIII Ly$\alpha$ &18.90                                         &18.97                  &   -             &24$^{+12}_{-10}$     &$\Delta$C=11\\  
\\
OVIII Ly$\alpha$(em) &18.99                                &18.97                  &4.70$^{+3.86}_{-3.32}$                &-23$^{+16}_{-18}$    &$\Delta$C=6 \\
\\
\hline
\multicolumn{6}{l}{(a) Errors are at the 90$\%$ confidence level for one interesting parameter.}\\
\multicolumn{6}{l}{(b) Improvement in the C-statistic (Cash 1979) for  adding a Gaussian line with two} \\
\multicolumn{6}{l}{     free parameters to a power-law model.}\\
\end{tabular}
\end{table*}

\section{RGS data analysis}
The first-order unbinned spectra from the two RGS cameras were simultaneously fitted in \emph{XSPEC} using the C-statistic (Cash 1979). 
The quoted errors correspond to 90$\%$ confidence level ($\Delta$C=2.71) for one interesting parameter.

The 6-35 $\AA$ (0.35-2 keV) continuum is well fitted by a single absorbed power-law with $\Gamma=2.25\pm 0.18$ (C=5264 for 4636 d.o.f.).  A quite steep power-law is not unexpected in this source known to have a strong soft excess.
We investigated whether an extended thermal component (APEC) is required by the data, as suggested by the \emph{Chandra} analysis. The fit did not improve significantly,  nor did the shape of the residuals (C=5261). Therefore we decided not to include any other continuum components. However, as discussed below, we checked for the presence of  weak collisional emission lines.

To start with a phenomenological approach, the spectra were divided into a number of regions containing about the same number of bins each (100). We performed an accurate inspection of every single region. Where the spectra showed a particularly evident absorption feature, a Gaussian profile with 0-width and a negative normalization  was added to the model. Only lines yielding an improvement in the fit higher than  $\Delta$C$>$5 (corresponding to $>$ 95$\%$ for two interesting parameters) were taken into account. For these structures the equivalent width was estimated and the corresponding atomic transition identified through several atomic databases such as CHIANTI (Dere et al. 1997; Landi et al. 2006), ATOMDB WebGUIDE \footnote{http://cxc.harvard.edu/atomdb/WebGUIDE/index.html} and NIST \footnote{http://physics.nist.gov}. The parameters of the most prominent lines are reported in Table~1 and the phenomenological fit is shown in Figure~1.  Two other absorption lines probably corresponding to FeXXI/MgXI and SIV/NVII transitions, although observed, are not reported in Table 1 because of their marginal detection  ($\Delta$C =5). None of the detected features is due to local (Galactic) absorption.

Finally we clearly detected an emission line ($\Delta$C=6) identified as OVIII Ly$\alpha$ (see Table~1). This is not an unusual result. Emission features are often observed in sources showing signatures of warm absorber (Blustin et al. 2002; Turner et al. 2004). 

Once the presence of absorption lines was ascertained via the phenomenological analysis, the next step was to study the physical properties of the gas using a self consistent modeling of the spectrum. Extensive simulations were carried out using  the photoionisation code XSTAR (v2.1ln9) \footnote{http://heasarc.nasa.gov/lheasoft/xstar/xstar.html}. The output was a grid of models (afterwards simply called XSTAR model) that can be fitted to the data in XSPEC. Each model in the grid represents a photoionised spectrum characterized by a particular value of the column density and the ionisation parameter. Our explored parameter range is N$_{H}$=[10$^{18}$, 10$^{23}$] and log$\xi$=[-3, 3]. We assumed a simple spherical geometry for the illuminated gas (covering factor =1), solar abundances (Grevesse \& Sauval 1998) , and an illuminating continuum from 1 to 1000 Rydbergs derived from the intrinsic spectral energy distribution SED of the source, with L$_{ion}\simeq 3\times 10^{45}$~erg~s$^{-1}$ (Figure~2). The turbulent velocity of the gas was set to v$_{turb}$=100~km~s$^{-1}$, as the limited RGS statistics did not allow to constrain the intrinsic width of the absorption features. 

The XSTAR model obtained from these simulations was applied as an \emph{XSPEC} multiplicative component to the power-law and fitted to the data. The fit significantly improved ($\Delta$C=160), as XSTAR can take into account any possible absorption feature in the spectrum. We note that the photon index is completely consistent  with the one obtained from a power-law only fit.
The presence of a highly ionised gas phase is confirmed, indeed the best fit required a high ionisation parameter log$\xi$=2.69$^{+0.05}_{-0.03}$~erg~cm~s$^{-1}$ and a column density N$_{H}$=3.19$^{+1.15}_{-0.79}$ $\times 10^{22}$~cm$^{-2}$. The robustness of these results  is attested by the N$_{H}$- log$\xi$ confidence contours (68$\%$ 90$\%$ 99$\%$) shown in Figure~3 (\emph{upper panel}). 
We  attempted to estimate the gas outflow velocities by measuring the line shifts between theoretical and observed values of the energy centroid  allowing the redshift parameter (z) of the XSTAR model to vary. We found reasonable outflow speeds  of the order of   -1200$^{+180}_{-200}$~km~s$^{-1}$, consistent with typical outflow velocities in Seyfert 1s (Blustin et al. 2002; Blustin et al. 2005) when the absorption gas is seen along the line-of-sight. Incidentally we point out that  the phenomenogical identifications proposed in Table~1 are confirmed by the  XSTAR model (see Figure~3 \emph{lower panel}).

We note that the energy centroid of the OVIII~Ly$\alpha$ emission line does not require any blueshift (although with large uncertainties). However it may suggest that,  contrary to the absorption lines observed along the line-of-sight,  the emission comes from an extended region around the source, maybe a shell, expanding in all directions.

\begin{figure}
\epsfig{file=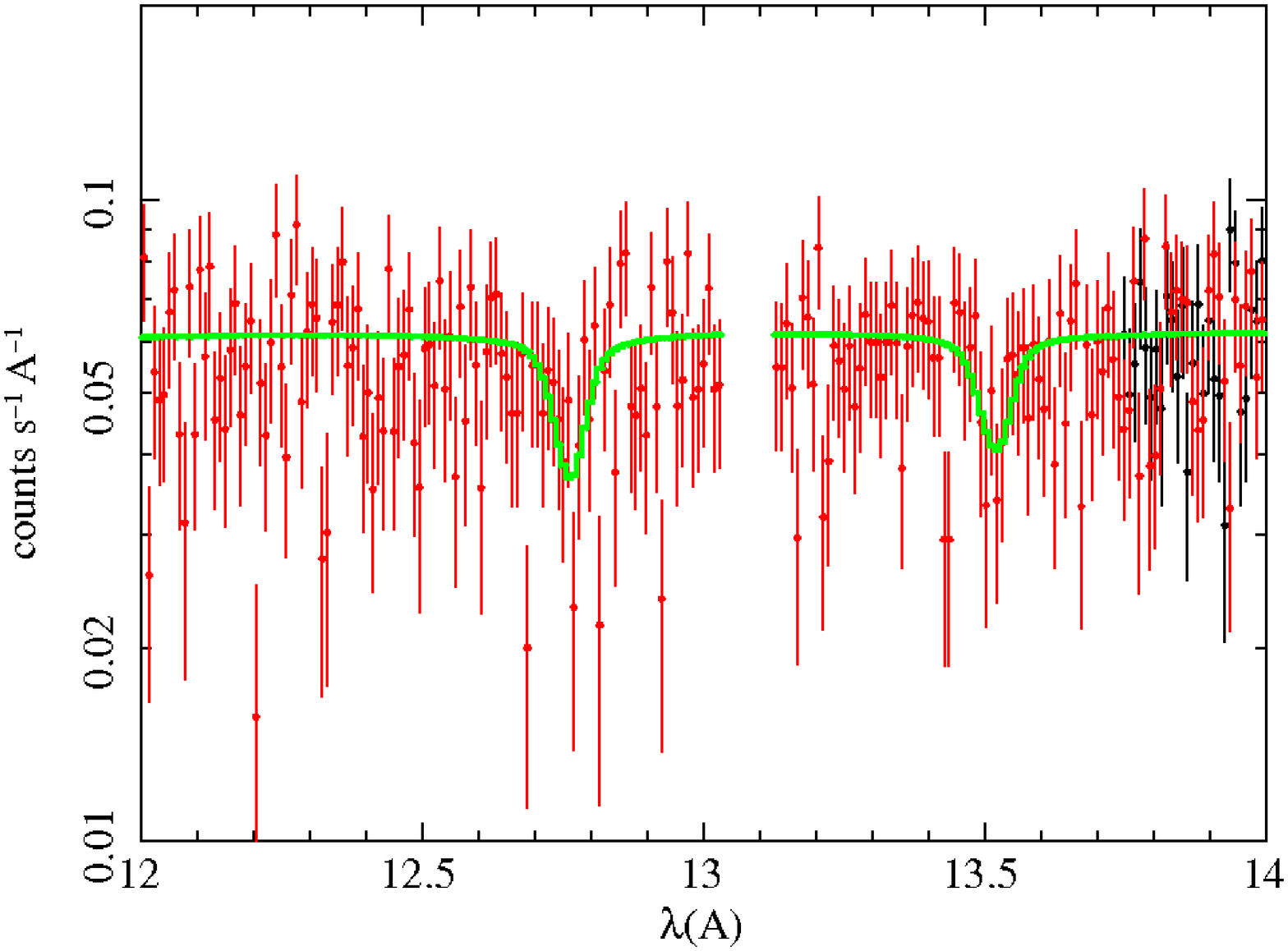,height=6.3cm,width=5.2cm, angle=-90}
\epsfig{file=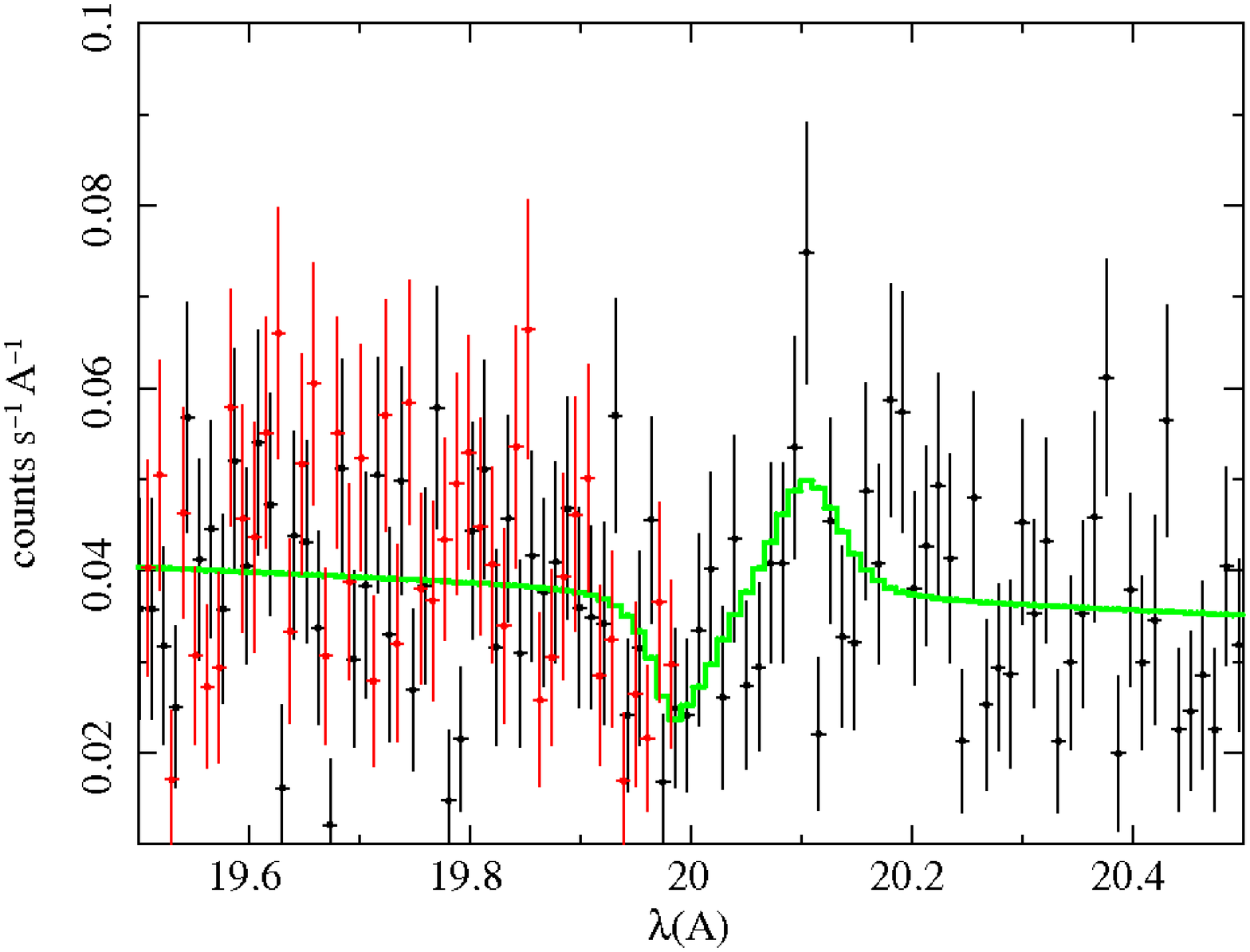,height=6.1cm,width=4.8cm, angle=-90}
 \caption{Zoom of the 3C~382 RGS observed-frame spectra (\emph{black}: RGS1; \emph{red}: RGS2), in wavelength ranges,  around the detected absorption and emission lines. The identified lines are: NeX-FeXX (\emph{upper panel}), OVIII Ly$\alpha$ in emission and absorption (\emph{lower panel}). The  \emph{green} line is the best fit phenomenological model (Table~1). }
   \label{fig1}
\end{figure}

\begin{figure}
\begin{center}
\epsfig{file =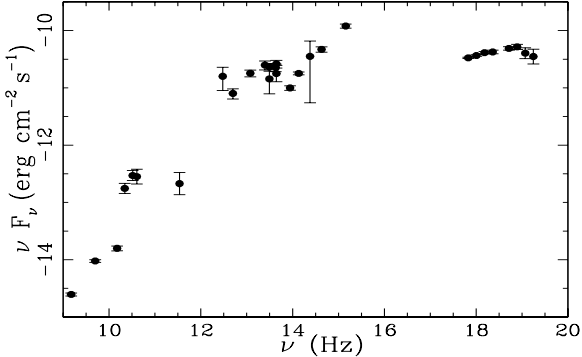,height=8cm,width=5cm, angle=-90}
    \caption{Radio to X-ray spectral energy distribution of 3C~382. Data are from literature (Rudnick et al. 1986; Knapp et al. 1990; Klavel et al.  2000; Lilly \& Longair 1982; Chiaberge et al. 2000; Chiaberge et al. 2002; Grandi et al. 2001; Quillen et al. 2003; Ramos Almeida et al. 2007; Hinshaw et al. 2007 ). Optical and UV measurements de-reddened with the extinction curve of Cardelli et al 1989. The visual extinction is deduced by the Galactic column density, assuming the gas-to-dust ratio from Shull et al. 1985} 
\label{fig2}
\end{center}
\end{figure}

\begin{figure}
\begin{center}
\epsfig{file=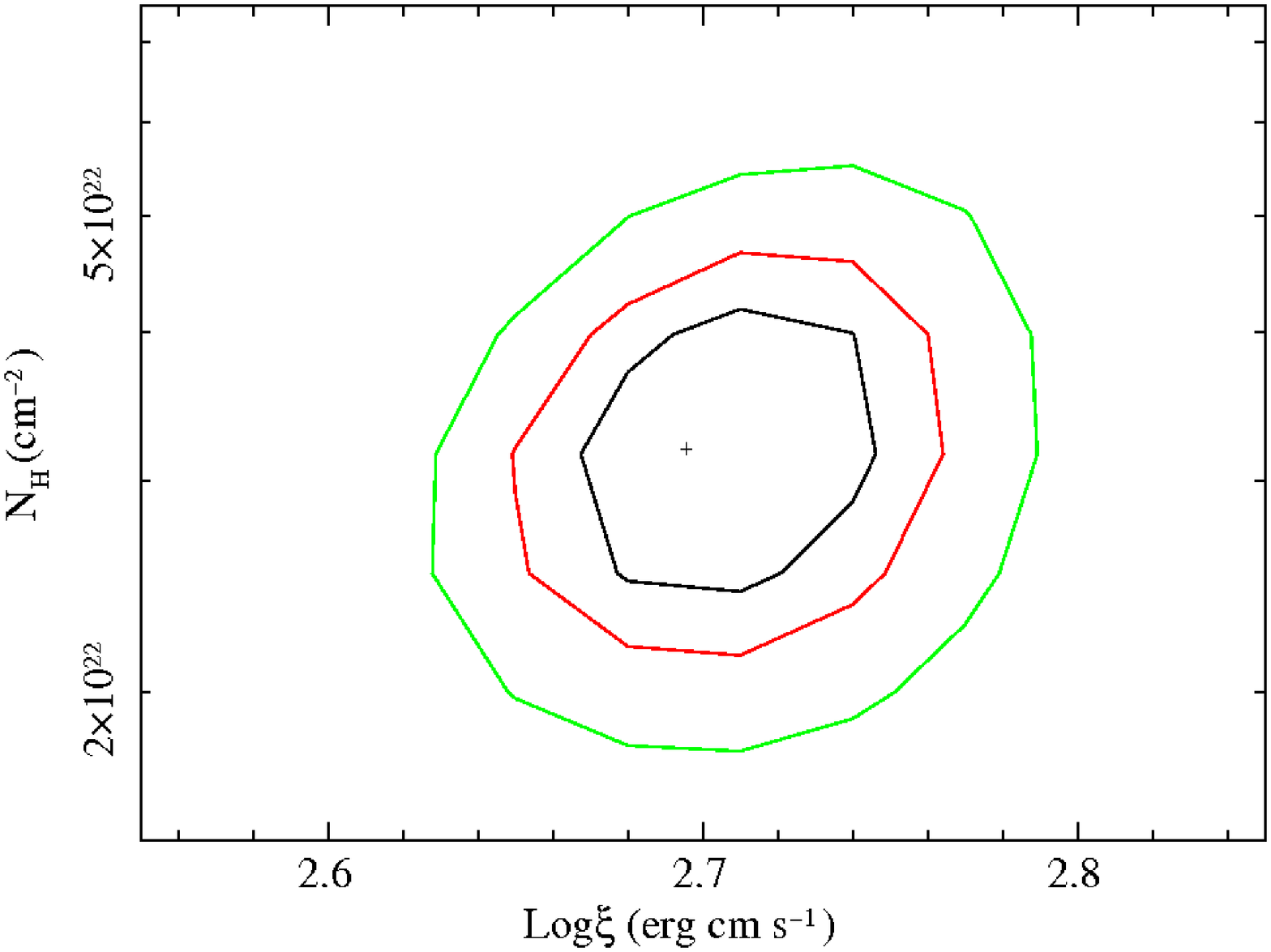,height=6cm,width=5cm, angle=-90} 
\epsfig{file=FIG3b.ps,height=6cm,width=5cm, angle=-90}
   \caption{\emph{Upper panel}: Contour plots for column density and ionisation parameter values of the warm absorber as obtained with the XSTAR model. The curves define the 68$\%$, 90$\%$ and 99$\%$ confidence intervals for two interesting parameters.  \emph{Lower panel}: Best fit rest-frame XSTAR model of 3C~382 in the overall RGS band (6-35 $ \AA$) corresponding to N$_{H}$=3.19$^{+1.15}_{-0.79}$ $\times 10^{22}$~cm$^{-2}$ and log$\xi$=2.69$^{+0.05}_{-0.03}$~erg~cm~s$^{-1}$.}
\label{fig3}
\end{center}
\end{figure}

\section{Discussion and conclusions}
The high-resolution analysis of the BLRG 3C~382 has revealed that relativistic collimated plasma and slower outflows of photoionised gas can coexist in the same object.
 The outflowing gas is found in a hot phase as attested by the derived ionisation parameter log$\xi$=2.69~erg~cm~s$^{-1}$ deduced with the XSTAR simulations.
 Most of the light elements are indeed completely ionised, and the strongest absorption features correspond to H-like elements and L-shell iron states. In Seyfert 1s warm absorbers are complex and multi-phase, often indicating that layers of gas at different ionisation states are part of the same
 outflow (Blustin et al. 2005). In the present case, only one component of hot gas was observed, without apparent traces of a colder component.
 Nonetheless, the physical conditions of the gas in 3C~382 warm absorber are quite similar to those measured in the highest ionised absorber  
in their radio-quiet counterparts (NGC~4051, Krongold et al. 2007; Mkr~279, Costantini et al. 2007; IRAS13349+2438, Sako et al. 2001). 
Taking into account the short exposure time and the very simple assumptions in the XSTAR model, the result appears encouraging. 
We are aware that setting v$_{turb}$ to an \emph{a priori} fixed value could potentially overpredict the column density if the real 
turbulent velocity is larger. Interestingly, also the EWs  of the absorption lines detected in 3C 382 are consistent with those generally measured 
in Seyfert 1s (Yaqoob et al. 2003; Turner et al. 2004). This similarity suggests a negligible jet contribution to the soft X-ray spectrum,
 although the large uncertainties in EW prevent more quantitative statements from being made.

In order to localize the warm absorber we estimate a minimum and a maximum distance for the gas following 
Blustin et al. (2005). The minimum distance is calculated with the assumption that the outflow velocity has to be greater or equal than the escape 
one, $R\geq \displaystyle\frac{2GM}{v_{out}^{2}}$.  For a M$_{BH}$=1.1$\times 10^{9}$ M$_{\odot}$ (Marchesini et al. 2004), 
and average outflow velocities of the order of 10$^{3}$ km~s$^{-1}$ (see Table 1) we deduce that the gas should be located at 
a distance larger than 10 pc. On the other hand, the maximum distance of the gas can be obtained by combining the ionisation 
parameter and the column density. If the gas is confined in a thin layer ($\Delta R /R \leq1$ with $\Delta R$ the layer depth), 
and the volume filling factor equal to 1, it turns out that $R\leq (\displaystyle\frac{L_{ion}}{\xi N_{H}})~\leq 60$~pc. 
According to Ghisellini \& Tavecchio (2008) the BLR and torus radii are expected to scale with the square root of the disk 
luminosity: $R_{BLR}=10^{17}$$L_{disk,45}^{1/2}$~cm and $R_{torus}=2.5 \times 10^{18}L_{disk,45}^{1/2}$~cm  (Cleary et al. 2007). 
Thus for 3C~382 we can estimate typical sizes for the BLR and dusty torus region, i.e. R$_{BLR}\simeq$0.06~pc and R$_{torus}\simeq$1.5~pc if L$_{disk}\sim L_{ion}$. 
These results suggest that the bulk of the highly ionised absorption gas, detected in 3C~382, is located beyond the torus, most likely in the NLR. 
Interestingly, we note that this is also the probable location of the photoionised emitting gas observed in 3C~33 (Torresi et al. 2009), 
another powerful RG seen at higher inclination angles. \\

\section*{Acknowledgments}
This paper is based on observations obtained with the \emph{XMM-Newton} satellite, an ESA funded mission with contributions by ESA Member States and NASA. We wish to thank the anonymous referee for valuable comments and detailed suggestions. We thank Massimo Cappi and Andy Pollock for very useful discussions. ET thanks for support the Italian Space Agency (contract ASI-I/088/06/0).

{}

\end{document}